\documentstyle[11pt,twoside,jltp]{article}

\title{Macroscopic Symmetry Group Describes Josephson Tunneling in
Twinned Crystals}

\author{M.B. Walker\address{Department of Physics, 
University of Toronto, Toronto, Ontario M5S 1A7, Canada} and 
J. Luettmer-Strathmann}

\runninghead{M.B. Walker and J. Luettmer-Strathmann}
{Macroscopic Symmetry Group}

\begin{document}

\begin{abstract}
A macroscopic symmetry group describing the superconducting state of
an orthorhombically twinned crystal of YBCO is introduced.  This
macroscopic symmetry group is different for different symmetries of
twin boundaries.  Josephson tunneling experiments
performed on twinned crystals of YBCO determine this macroscopic 
symmetry group and hence determine the twin boundary symmetry (but
do not experimentally determine whether the microscopic order
parameter is primarily d- or s-wave). A consequence of the
odd-symmetry twin boundaries in YBCO is the stability of
$\frac{1}{2} \Phi_0$ vortices at the intersection of a twin boundary
and certain grain boundaries.
\end{abstract}

\maketitle


\section{INTRODUCTION}
Measurements of Josephson tunneling represent an important method of
determining the symmetry of the superconducting state of high
temperature superconductors\protect\cite{bra94,igu94,mat95,mil95,sun94,
tsu96,wol93}.  This
article presents, in a 
pictorial fashion, symmetry arguments which give insight into
precisely what is being measured experimentally in experiments
performed on twinned crystals.  (Many experiments on 
YBa$_2$Cu$_3$O$_{7-x}$ (YBCO) have been carried out on 
twinned crystals\protect\cite{igu94,mat95,mil95,sun94,
tsu96,wol93}.  Interesting studies of these twins can be
found in\protect\cite{ame90,zhu93}).  The important concept 
introduced here is that of a 
macroscopic symmetry group describing the symmetry properties of the 
superconducting state of the twinned crystal as a whole.  This
macroscopic symmetry group is what is determined by Josephson 
experiments on highly twinned single crystals of YBCO. The macroscopic 
symmetry group of the twinned crystal depends not only on the
microscopic symmetry group of a single twin (from which it differs)
but also on the symmetry of the twin boundaries.  Current
experimental results indicate that the macroscopic symmetry group
describing the superconducting state of twinned YBCO is
D$_4^{(1)}$(D$_2$)$\times$I$\times$R in the notation 
of\protect\cite{vol85};
this is the same symmetry group that describes $d_{x^2-y^2}$
superconductivity in a tetragonal superconductor.  This result
then leads to the conclusion that the twin boundaries in YBCO
have odd reflection symmetry (and says nothing about whether
the microscopic state of a single twin is predominantly s or
d wave).  Similar conclusions were reached in 
\protect\cite{wal96a} by twin averaging the Josephson currents (but
without introducing the idea of macroscopic symmetry).  The
macroscopic symmetry of a twinned sample for the case where the
superconducting state of a twin breaks time-reversal symmetry is
studied in\protect\cite{wal96b}.

An interesting consequence of the odd symmetry of the twin boundaries
in YBCO is the predicted\protect\cite{wal96c}
existence (even in the absence of external
magnetic fields) of stable vortices carrying a flux of
$\frac{1}{2} \Phi_0$ at the intersection of a twin boundary and
an asymmetric 45$^{\circ}$ grain boundary. This
forms the basis for an explanation of the spontaneously generated
magnetic flux observed by
Mannhart et al.\protect\cite{man96} and Moler et al.\protect\cite{mol96}
in 45$^{\circ}$ grain boundaries.

\section{MACROSCOPIC SYMMETRY OF YBCO}
This paper assumes that the superconductor YBCO is orthorhombic and
has a second order phase transition to its superconducting state.
According to the Ginzburg-Landau theory of phase transitions there
are therefore eight possible symmetries for the superconducting order
parameter, one corresponding to each irreducible representation
of the orthorhombic point group.  The fact that Josephson tunneling has
been observed to occur along all three principal orthorhombic
directions from YBCO into an isotropic superconductor such as lead
eliminates\protect\cite{wal96b} 
all of these possibilities except one, namely an order
parameter of A$_{1g}$ (or $ux^2 + vy^2$) symmetry (here $u$ and $v$
are arbitrary constants).  Since $ux^2+vy^2 = d(x^2-y^2) + s(x^2+y^2)$,
where the constants $d$ and $s$ can be determined in terms of 
$u$ and $v$, the order parameter can also be viewed as being a
linear combination of the tetragonal symmetry types $x^2-y^2$
(i.e. d wave) and $x^2+y^2$ (i.e. s wave).

Within the framework of a Ginzburg-Landau model for the twin
boundary, the superconducting state must be either even or odd
with respect to a reflection in a twin 
boundary;\protect\cite{wal96a,and87}  which
case actually occurs (i.e. which has the lower free energy) depends 
on the detailed microscopic structure and physics of the superconducting
state at the twin boundary, about which this article makes no a priori
assumptions.  The two types of twin boundaries are illustrated 
schematically in Fig.~\ref{fig1}.  This figure shows two equivalent 
representations of the microscopic symmetry of the superconductivity
in the two different types of twins.  On the right of Fig. 1(a) and (b)
the ellipses represent the orthorhombic $ux^2+vy^2$ symmetry;  the
equivalent representation on the left shows the separate
tetragonal $x^2-y^2$ (the four-lobed objects) and $x^2+y^2$
(the circles) components.  Notice that for the odd-symmetry
twin boundary, all of the plusses and minuses change sign on a 
reflection in the twin boundary, whereas for the even-symmetry
twin boundary, the superconducting state is invariant with
respect to a reflection in the twin boundary.

\begin{figure}
%
%
\makebox[5in]{
\includegraphics{mosfig1.ps}
\rule[0.8in]{0in}{0.8in}
}
\caption{Even and odd symmetry twin boundaries.  The $c$
axis of the YBCO film is normal to the page and the
orientation of the crystallographic $a$ and $b$ axes in
the two twins are as shown on the right of (a).}
\label{fig1} \end{figure}

In a twinned crystal there are two orientations of the
twinning planes, namely the (110) and the $(\bar{1}10)$ twinning 
planes.  Fig. 2 shows regions of (110) and $(\bar{1}10)$ twinning
planes and their interface, as observed in 
experiment.\protect\cite{zhu93}
If the twin boundaries have odd symmetry, then the phase of
the order parameter will be propagated in a particular way
throughout the crystal by changing sign at each twin boundary,
as is seen from Fig. 2(a).  Similarly, if the twin
boundaries have even reflection symmetry, there is no change of
sign of the order parameter at the twin boundaries, as is the case 
in Fig. 2(b).

\begin{figure}
%
%
\makebox[5in]{
\includegraphics{mosfig2.ps}
\rule[1.125in]{0in}{1.125in}
}
\caption{Macroscopic symmetry for odd and even twin boundaries.}
\label{fig2} \end{figure}

A remarkable feature of Fig. 2 is that, even though the
microscopic superconducting state of the individual twins is
assumed to be the same in both cases, (a) and (b) have very 
different appearances as a result of the fact that the twin
boundaries are different in the two cases.  In fact (a) and (b)
have different macroscopic symmetries.  To determine the
macroscopic symmetry, the region of the crystal shown
in (a) must be imagined to be part of a large crystal in
which regions of (110) and $(\bar{1}10)$ twin boundaries are
randomly distributed.  Now if the crystal of Fig. 2(a)
is rotated by 90$^{\circ}$ about the $c$ axis (i.e. an axis
normal to the page), it will look the same except for a 
change of sign of all the plusses and minuses.  This overall
change of sign under a rotation of 90$^{\circ}$ about the $c$ axis
is one of the characteristics of $d_{x^2-y^2}$ superconductivity
in a tetragonal superconductor.  More detailed study shows that
a twinned sample with odd-symmetry twin boundaries has exactly
the same symmetry elements in its macroscopic symmetry group as does the
(microscopic) symmetry group of a tetragonal $d_{x^2-y^2}$
superconductor; this group is denoted\protect\cite{vol85} by
D$_4^{(1)}$(D$_2$)$\times$I$\times$R. 
Similarly, the crystal of Fig. 2(b) is invariant
under a rotation of 90$^{\circ}$ about the $c$ axis, and furthermore
has precisely the same elements in its macroscopic symmetry group as
are found in the (microscopic) symmetry group of a tetragonal s-wave
superconductor; this group is called\protect\cite{vol85}
D$_4 \times$I$\times$R.

There is substantial evidence from Josephson 
experiments\protect\cite{igu94,mat95,tsu96,wol93} 
that the macroscopic symmetry of twinned YBCO is that of a 
$d_{x^2-y^2}$ tetragonal superconductor.  The appropriate conclusion 
to draw from this result is that the twin boundaries in YBCO 
have odd reflection symmetry.

The macroscopic symmetry group associated with various
microscopic superconducting states which break 
time-reversal symmetry can also be determined.  For
example, the macroscopic symmetry group of the state
described in \protect\cite{sig96} is the same as
that of a tetragonal $d + is $ superconductor.

The main point of this section was to try to demonstrate pictorially
and convincingly how the macroscopic symmetry of a twinned crystal is 
determined by the symmetry of the twin boundaries.  

\section{ASYMMETRIC 45$^{\circ}$ GRAIN BOUNDARIES}
Fig. 3 shows what is called\protect\cite{man96,mol96} an
asymmetric 45$^{\circ}$ grain boundary in a YBCO film.  The
crystal on the left consists of two twins, and the symmetries
of the superconducting order parameters in the three regions
of the figure are represented by ellipses, as described above.
The grain boundary acts as a Josephson junction with a free energy
per unit area at point y (y is distance measured along the
grain boundary) given by
$$
F = -\frac{\Phi_0 j_c}{2\pi c} cos[\gamma (y) - \alpha (y)],
$$
where $j_c$ is the critical current density,  and $\gamma$ is the
gauge invariant phase difference of the order parameter across
the grain boundary.  This is the usual expression except for
$\alpha (y)$, which is zero between 0 and q (see Fig. 3) but is
$\pi$ between p and 0 to take account of the change of sign at
the twin boundary of the superconducting order parameter 
to the left of the grain boundary.  To minimize this free
energy, $\gamma (y)$ will be $\pi$ from p to 0, will vary from
$\pi$ to zero at the twin boundary (changes in $\gamma$ take
place over a length $\lambda_J$, the Josephson penetration
depth) and be zero in the region from 0 to q.  Variations in 
$\gamma$ are associated with the presence of a magnetic field
in the junction\protect\cite{jos65}, and the variation of 
$\gamma$ by $\pi$ at the twin boundary corresponds to a vortex 
there with flux $\frac{1}{2} \Phi_0$.  The stability of this
$\frac{1}{2} \Phi_0$ vortex is a consequence of the odd
reflection symmetry of the twin boundary.  Further details 
of the properties of asymmetric 45$^{\circ}$ grain boundaries
can be found elsewhere \protect\cite{wal96c,man96,mol96}.

\begin{figure}
%
%
\makebox[5in]{
\includegraphics{mosfig3.ps}
\rule[1.125in]{0in}{1.125in}
}
\caption{Asymmetric 45$^{\circ}$ grain boundary.}
\label{fig3} \end{figure} 

\section{CONCLUSIONS}
The relationship of the twin boundary symmetry to the 
macroscopic symmetry measured by Josephson tunneling on
an orthorhombically twinned superconductor such as YBCO
has been demonstrated pictorially.  An interesting
consequence of the odd-symmetry twin boundaries found in
YBCO is the existence of a
$\frac{1}{2} \Phi_0$ vortex at the intersection of an 
isolated twin boundary and an asymmetric 45$^{\circ}$ grain
boundary.

\section*{ACKNOWLEDGMENTS}
We thank the Natural Sciences and Engineering Research Council
of Canada for its support.


\begin{thebibliography}{99}

\bibitem{bra94}
D.~A. Brawner and H.~R. Ott, {\it Phys. Rev. B} {\bf 50},  6530  (1994).

\bibitem{igu94}
I. Iguchi and Z. Wen, {\it Phys. Rev. B} {\bf 49},  12388  (1994).

\bibitem{mat95}
A. Mathai, Y. Gim, R.~C. Black, A. Amar, and F.~C. Wellstood, 
{\it Phys. Rev. Lett.} {\bf 74},  4523  (1995).

\bibitem{mil95}
J.~H. Miller, Jr., Q.~Y. Ying, Z.~G. Zou, N.~Q. Fan, J.~H. Xu,
M.~F. Davis, and J.~C. Wolfe, {\it Phys. Rev. Lett.} {\bf 74},
2347 (1995).

\bibitem{sun94}
A.~G. Sun, D.~A. Gajewski, M.~B. Maple, and R.~C. Dynes, {\it Phys. 
Rev. Lett.} {\bf 72},  2267  (1994).

\bibitem{tsu96}
C.~C. Tsuei, J.~R. Kirtley, M. Rupp, J.~Z. Sun, A. Gupta, and M.~B. Ketchen, 
{\it Science} {\bf 271},  329  (1996).

\bibitem{wol93}
D.~A. Wollman, D.~J. Van Harlingen,
W.~C. Lee, D.~M. Ginsberg, and A.~J. Leggett, 
{\it Phys. Rev. Lett.} {\bf 71},  2134  (1993).

\bibitem{ame90}
S. Amelinckx, G. Van Tendeloo and J. Van Landuyt,
{\it Solid State Ionics} {\bf 39}, 37 (1990).

\bibitem{zhu93}
Y. Zhu, M. Suenaga, and J. Tafto, {\it Philosophical Magazine A}
{\bf 67}, 1057 (1993).

\bibitem{vol85}
G.~E. Volovik and L.~P. Gor'kov, {\it Sov. Phys. JETP} 
{\bf 61},  843  (1985).

\bibitem{wal96a}
M.~B. Walker, {\it Phys. Rev. B} {\bf 53}, 5835 (1996).

\bibitem{wal96b} 
M.~B. Walker and J. Luettmer-Strathmann, to be published in
{\it Phys. Rev. B}, July 1, 1996.

\bibitem{wal96c}
M.~B. Walker, submitted to {\it Phys. Rev. B}.

\bibitem{man96}J. Mannhart, H. Hilgenkamp, B. Mayer, Ch. Gerber,
J.~R. Kirtley, K.~A. Moler and M. Sigrist, submitted to {\it Phys. Rev. Lett.}

\bibitem{mol96}K.~A. Moler, J.~R. Kirtley, J. Mannhart,
H. Hilgenkamp,
B. Mayer, Ch. Gerber, Ruixing Liang, Douglas~A. Bonn and
Walter~N. Hardy, to be published in {\it Conference Proceedings of
10$^{th}$ Anniversary HTS Workshop on Physics, Materials and
Applications, Houston, Texas} (1996).

\bibitem{sig96} M. Sigrist, K. Kuboki, P.~A. Lee, A.~J. Millis,
and T.M. Rice, {\it Phys. Rev. B} {\bf 53}, 2835 (1996).


\bibitem{and87}
A.~F. Andreev, {\it JETP Lett.} {\bf 46},  584  (1987).

\bibitem{jos65}
B.~D. Josephson, {\it Adv. Phys.} {\bf 14}, 419 (1965).

\end{thebibliography}
\end{document}